\documentstyle[aps,prb,multicol,psfig]{revtex}
\begin{document}
\title{Stripe formation in  electron-doped cuprates}
\author{A. Sadori and  M. Grilli}
\address{Istituto Nazionale di Fisica della Materia and
Dipartimento di Fisica, Universit\`a di Roma ``La Sapienza'',\\
Piazzale A. Moro 2, Roma, Italy 00185}
\maketitle

\begin{abstract}
We investigate the formation of charge domain walls in an electron-doped
extended Hubbard model for the superconducting cuprates. 
Within an unrestricted Hartree-Fock approach, extended by the
introduction of slave-bosons to obtain a more proper treatment 
of strong correlations, we demonstrate the occurrence of 
stripes in the (1,1) and (1,-1) directions having one doped electron
per stripe site. The different filling, direction and width of these
electron-doped stripes with respect to those obtained in the hole-doped
systems have interesting observable consequences, which are discussed. 
\end{abstract}


\begin{multicols}{2}
The issue of charge inhomogeneity in the cuprates and particularly in the
underdoped region of their phase diagram is presently attracting
a great attention both from the theoretical and the experimental 
point of view \cite{STRIPES98}. 

Some theoretical proposals 
are based on the occurrence of stripe phases
with the charge tending to order in incommensurate charge-density
waves or in fluctuating domain walls
to explain both the anomalous normal state
properties and superconductivity
\cite{CDG,EMERY1,ZAANEN}. In particular it was proposed
\cite{CDG} that the stripe phase in the underdoped cuprates 
is ruled by a charge-driven critical line ending in a 
hidden quantum critical point (QCP)
located around optimal doping. Although this QCP may be masked by 
(strong) pairing due to the nearly singular attractive
charge fluctuations \cite{CDGZP,CAPECOD} the physics of the 
underdoped cuprates is strongly influenced by the tendency
to spatial charge order. 

On the experimental side, the 
critical line for stripe formation has likely been detected
in NQR \cite{IMAI} and 
in magnetotransport experiments \cite{BOEBINGER,FOURNIER}, 
where the suppression of superconductivity by strong magnetic fields 
both in hole- and electron-doped materials reveal a metal-to-insulator
crossover line ending at $T=0$ near optimal doping. 
In the hole-doped compounds the insulating
character of the low-temperature and low-doping phase is stronger
at the specific
doping 1/8, where commensurability effects between the charge ordering
and the underlying lattice make the charge order particularly 
robust \cite{CAPECOD}. This latter effect is a signature of spatial
order and provides further evidence of the enhanced strenght
of stripe ordering in various compounds at specific fillings
(this is the so-called 1/8 phenomenon). This commensurability
effect seems instead much less evident in the electron-doped materials.

There is also an increasing and more direct experimental evidence 
for the occurrence of (static or dynamic) charge ordering
in under- and optimally doped cuprates \cite{STRIPES98}, ranging from
EXAFS \cite{BIANC} to neutron scattering \cite{TRAN}, and 
X-ray scattering \cite{BIANX,NIEMOELLER}.

The stripe instability was predicted theoretically in Ref. 
\onlinecite{ZAAGUNN}
within Hartree-Fock (HF) theory applied to the extended Hubbard model
and confirmed by a number of subsequent investigations \cite{VARIE}.
Only recently, however, the occurrence in the single-band Hubbard model 
of half-filled stripes (i.e. with half a hole per stripe site) in the (1,0) [or
equivalently in the (0,1)] direction 
[Half-Filled-Vertical stripes (HFVS)] was demonstrated
within a generalized HF approach
including slave bosons to suitably handle the strong local
electron-electron interaction. This achievement remarkably
showed that (a proper treatment of) 
strong correlations and long-range interactions
are crucial in determining a specific  stripe texture
similar to the one observed in the cuprates. 
 
The aim of the present work is to apply the same joynt 
HF--slave-boson technique to
theoretically investigate the 
stripe formation in a more realistic extended three-band Hubbard model
in the case of electronic doping. It is to be borne in mind that in
any HF approach the solutions are  self-consistently
stabilized by locating the chemical potential in the middle of
a gap. Therefore this approach is only suitable to describe
insulating phases where the filling and/or the breaking
of translational symmetry gives rise to such gaps. In our case
we therefore intend to describe by our method the 1/8 commensurate systems
where both the charge ordering and the (nearly) insulating
character are substantial.

{\em ---The model and the main outcomes.}
The starting point is the extended three-band Hubbard model

\begin{eqnarray}
H&=&\sum_{<ij>\sigma}t^{pd}_{ij}(d^\dag _{i\sigma}p_{j\sigma} + h.c.)
+\sum_{<jj^\prime>\sigma}t^{pp}_{jj'}(p^\dag _{j\sigma} p_{j'\sigma} + h.c.)
\nonumber\\
&+&\Delta\sum_{j\sigma}p^\dag _{j\sigma}p_{j\sigma}
+U_d\sum_i d^\dag _{i\uparrow} 
d_{i\uparrow} d^\dag _{i\downarrow} d_{i\downarrow} \nonumber \\
&+&U_p\sum_j p^\dag _{j\uparrow} 
p_{j\uparrow} p^\dag _{j\downarrow} p_{j\downarrow}
+\sum_{i\neq j,\sigma\sigma^\prime}
V_{ij}c^\dag _{i\sigma} c_{i\sigma} c^\dag _{j\sigma^\prime}
 c_{j\sigma^\prime}
\label{model}
\end{eqnarray}
where the first and second terms describe the  nearest neighbor (nn)
copper-oxygen and oxygen-oxygen hopping of holes
in the $d$ copper orbitals and in the $p$ oxygen orbitals respectively.
$\Delta$ is the energy difference between $p$ and $d$ orbitals  (the energy
of the $d$ orbitals is set to zero), while $U_d$ $(U_p)$ is the
local Hubbard repulsion on the copper (oxygen) orbitals. The last
term is the long-range interaction, with 
$V_{ij}=Ve^{-r_{ij}/\lambda_S}/\left(2r_{ij}\right)$, 
with $r_{ij}$ being the distance between two sites, 
$\lambda_S$ is the screening length. $V$ is set to 
$V=U_{pd}e^{1/\l_S}$ in order to match the $U_{pd}$ nn  
 Cu-O repulsion. In this term, the $c$ operators represent 
$d$ or $p$ hole operators depending on the nature of the
site.

Following the procedure of Ref.\onlinecite{SCDG} we treat the strong
local repulsion $U_d$ by means of a slave-boson approach.
The weaker strength of the repulsion on $p$ orbitals and
the smaller average hole density on oxygen sytes makes
unnecessary to apply the same technique to the oxygen orbitals.
Therefore the other interaction terms $U_p$ and $V$ are treated
by a standard HF decoupling. 

We summarize here the main results valid both for hole and
electron doping. i) Like in the 
single-band Hubbard model, also in the three-band Hubbard
model the major effect of the slave-boson treatment is to
stabilize the stripes \cite{GOETZ} with respect to isolated 
spin polarons. This stability mostly arises because, according to general
arguments \cite{OLESZAANEN}, the charge mobility tends to favor
stripes with respect to isolated spin polarons. Then, since
the Gutzwiller-like procedure implemented via the slave bosons
allows the charge to delocalize without paying a much too large
repulsive energy stripes are not artificially 
disfavored within our approach.
The same is not true in standard HF treatments,
where the double occupation of the sites is avoided via a strong
antiferromagnetic polarization of the spins, which greatly suppresses
the intersite hopping. ii) For moderate to large values of
$U_d$, the natural energy scale for charge excitations 
no longer is $U_d$, but rather
the charge-transfer energy $\Delta$ becomes the principal
parameter determining the stability of stripes with respect to
an ordered lattice (Wigner crystal) of maximally
separated spin polarons. The stripes are only
stable below a critical value $\Delta_c$, depending on the
other parameters of the model, which in any reasonable realistic case 
is much larger than the critical values obtained  in the
standard (i.e. without slave-bosons) HF technique. 
iii) For a quite broad range of realistic values
of the parameters the HFVS and the
Filled Diagonal stripes [i.e. with one hole per stripe site and 
running along the (1,1) or (1,-1) directions (FDS)]  are the most
stable or the only stable solutions. In particular, {\em for hole
doping} $x=1/8$, while the 
standard HF treatment favors the FDS, it was remarkably found that
{\em the HFVS become the ground state when strong correlations are
properly treated by slave bosons}. The absolute
stability of HFVS occurs even in the absence of
long-range interactions. In this respect, the 
hole-doped three-band Hubbard model seems more favorable than
the single-band one in giving rise to half-filled stripe textures of the
type observed in cuprates. iv) The local repulsion on oxygen only
plays a minor role, while v) the oxygen-oxygen hopping $t_{pp}$ 
enhances the charge mobility and therefore
stabilizes the stripes with respect to
the polaron Wigner crystal (POLWC). As far as the effects of the ``long-range''
interaction are concerned, it was found that a nn repulsion
$U_{pd}$ naturally disfavors the stripes while it does not affect
the POLWC, where the charges are at
larger distance. Nevertheless vi) a  moderate $U_{pd}$ stabilizes the HFVS
with respect to the FVS, thus strengthening the stability of this
configuration. On the other hand, vii) the large-distance part of
a truely long range interaction turns out to increase the energy of
the POLWC thus favoring further the stability of an array of stripes. 

Finally, the effects of a coupling between holes and static
lattice deformations were also investigated. In particular, 
the hopping was modulated by the breating displacement 
of oxygen ions in the $j$ site 
$t_{ij}(\{u_j\})=t_{pd}\pm\alpha u_j$, where the sign is positive
if the Cu-O distance is decreased, while it is negative
if the Cu-O distance increases. For simplicity, in this model we neglected 
the coupling between the O-O hopping and the lattice.
In an analogous way the orbital energy on the copper site is
modified as $\epsilon_i(\{u_k\})=\epsilon_d+\beta\sum_j (\pm u_j)$.
Then the dimensionless hole-lattice coplings are
 $\lambda_\alpha=\alpha^2/(Kt_{pd})$ and
$\lambda_\beta=\beta^2/(Kt_{pd})$.
At variance with the findings of Ref.\onlinecite{BISHOP},
where stripe textures were only found as {\em local} minima, the 
slave-boson treatment produces absolutely stable stripe solutions.
However, owing to the specific choice of lattice displacement,
there occurs a frustration effect whenever holes (or electrons) 
are located on neighboring sites. This frustration of the
lattice displacement is not present in the case of distant 
polarons. Therefore 
in stripe configurations (particularly in the filled ones), viii)
the electron-phonon coupling turns out to be detrimental for
stripes in comparison to the POLWC. Nevertheless, 
the absolute stability of HFVS at hole-doping $x=1/8$
survives in the presence of a moderate coupling. 

For concreteness, we report in Table I the energy difference
between the HFVS or the FDS and the POLWC configurations at $x=1/8$
for a typical set of parameters in a $9\times 8$ lattice. 
The HFVS stabilization effect of
the long-range (LR) interaction is shown in the second
line, while the destabilization due to the electron-phonon (EP)
coupling is shown in the third line. The consideration of 
all the different interactions leads to a stable HFVS configuration
as indicated in the fourth line. As shown in the last column,
the FDS are instead always disfavored with respect to the 
POLWC. 
\vspace {.5 truecm}

\begin{tabular}{l|c|c}
& $\epsilon_{HFVS} -\epsilon_{POLWC}(eV)$& 
$\epsilon_{FDS} -\epsilon_{POLWC}(eV)$\\ \hline
HSC     & $-0.036$                  &      $+0.003$         \\
LR      & $-0.084$                  &          ?          \\ 
EP      & $+0.009$                  &      $+0.034$       \\ 
EP+LR   & $-0.034$                  &          ?       \\
\end{tabular}
\vspace {0.5 truecm}

{\small Table I: Energy differences (per doped hole)
between two different stripe configurations
(HFVS and FDS) and the POLWC in a $9\times 8$ (or $8\times 8$)
lattice with periodic
boundary conditions. According to Ref. \onlinecite{HSC},
the parameter values are $t_{pd}=1.3 eV$, $t_{pp}=0.65eV$,
$U_d=10.5eV$, $U_p=4eV$. The LR coupling is such that $U_{pd}=1.2eV$
with a screening length $\lambda_S=32$ lattice units. The EP
couplings are $\lambda_\alpha=0.5$ and $\lambda_\beta=0$.}
\vspace {.5 truecm}

{\em --- Electron doping.} While the above analysis was mostly carried out for 
hole doping (and particularly for $x=1/8$), it is quite 
important to extend the investigation to electron-doped systems.
While the general validity of the above
results is mantained, some important differences are also found. Specifically
it was noticed that the charge in electron doped stripes is more localized.
A comparison of the charge profile of hole-doped (HD) and electron-doped
(ED) stripes is shown in Fig. 1, where the narrower charge distribution
of ED stripes is apparent.
\begin{figure}
{\hspace{0.3cm}\vspace{-.5cm}
{\psfig{figure=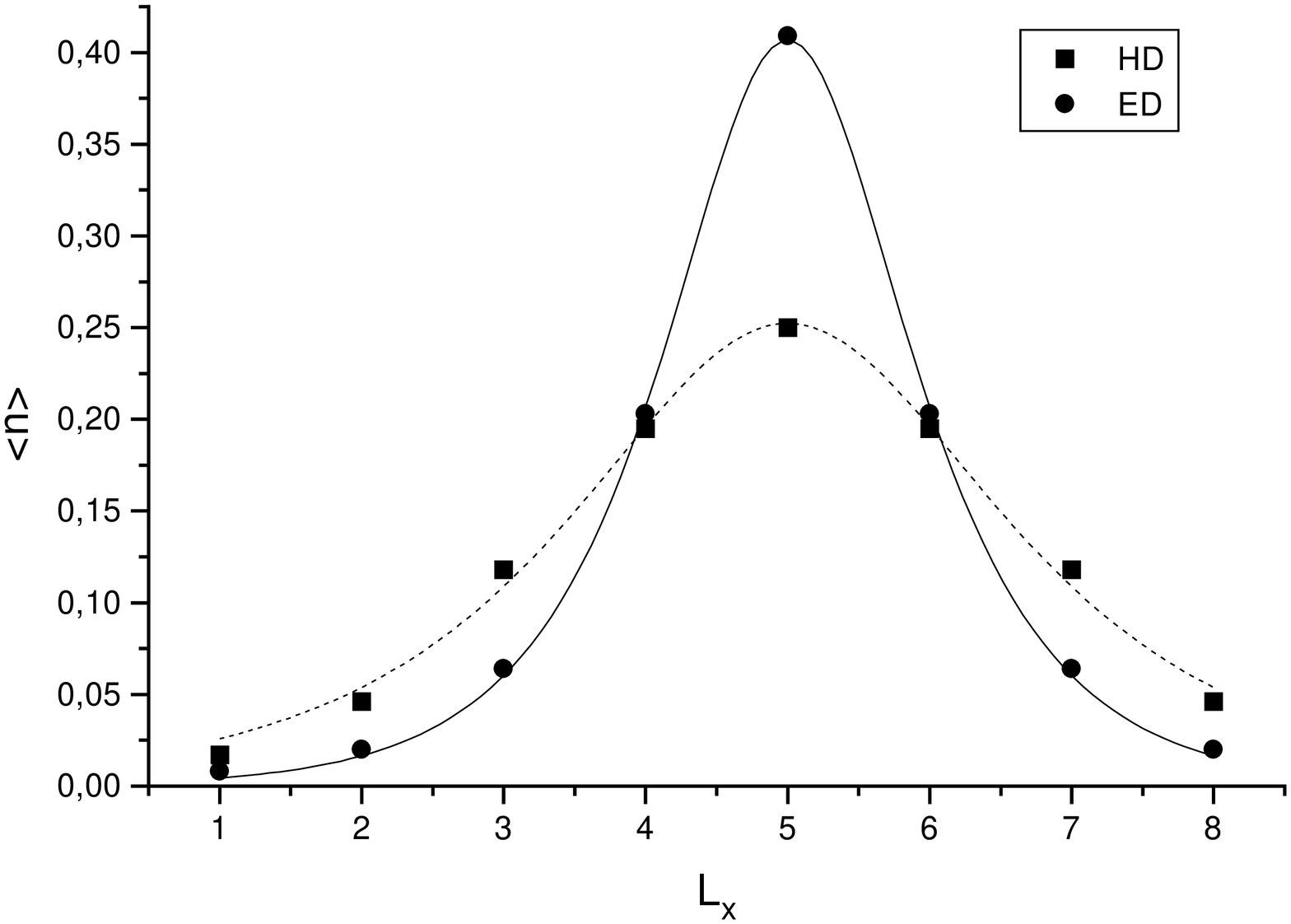,width=6.0cm,angle=0}}}
\end{figure}
\begin{figure}
{\hspace{0.3cm}
{\psfig{figure=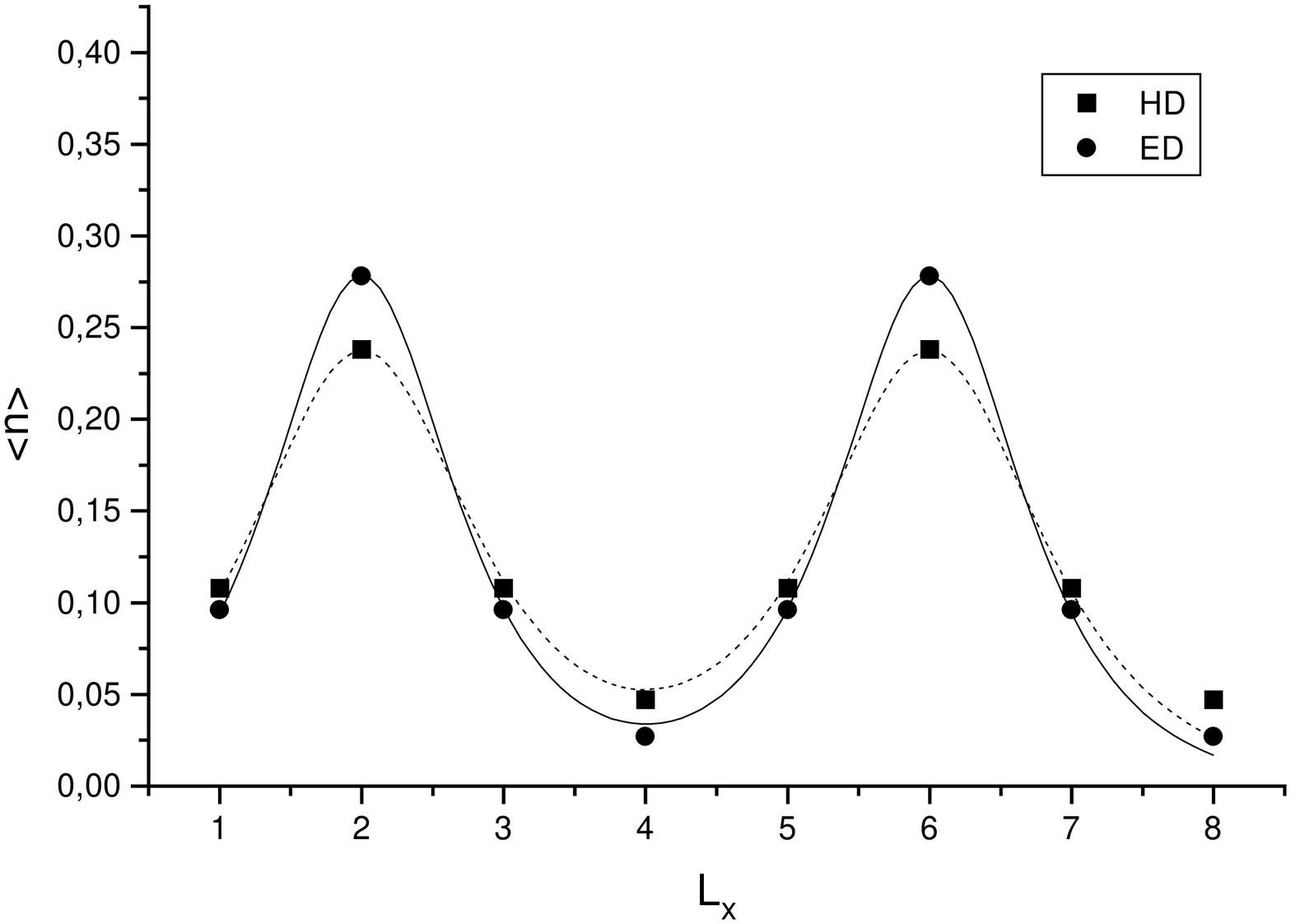,width=6.0cm,angle=0}}}
\end{figure}
\vspace{.5cm}
{\small FIG. 1. Charge density distribution in the 
(1,0) direction for FDS (i.e. in the (1,1) direction, above)
and for HFVS (i.e. in the (0,1) direction, below) for both
ED and HD systems. The parameters are the same as in Table I
with $x=\pm 1/8$, but with $\lambda_\alpha=\lambda_\beta=0$.}
\vspace{.5cm}

This effect is quite natural since the doped holes mostly
reside on oxygen sites where due the smaller interaction and to 
the direct $t_{pp}$ hopping acquire a high mobility, while the
doped electrons mostly reside on copper sites. Therefore the
doped-electron mobility is lower because it only occurs via an
intermediate  charge transfer to oxygen sites. The smaller
mobility and the consequent relative narrowness of the stripes
in ED systems has two main consequences. The first is that the
stripe formation is generically more difficult. Nevertheless
stripe textures still remain the ground state for realistic
values of the parameters. In particular, even in the presence
of a sizable electron-lattice coupling (which, as discussed above, 
favors the POLWC solution more than the stripes) stable FDS
textures are found. A second important effect is that
{\em FDS (with one doped electron per stripe site) 
become more stable than the HFVS}. This latter finding
suggests that ED cuprates should be more similar to the nickelates,
where diagonal stripes with one doped hole per site are detected.

{\em --- Discussion.}
The above reported findings have important consequences.
First of all, the fact that FDS accomodate one doped charge
per site, rather than one-half like the HFVS, increases the
(average) distance between the stripes. For instance, at
$|x|=1/8$, in the FDS case the stripes are at distance
$d_{FDS}=8a/\sqrt{2}$, while the HFVS are separated by
$d_{HFVS}=4a$. As a consequence the interstripe antiferromagnetically
correlated background is better preserved when FDS are formed.
This effect is emphasized by the narrow width of the ED stripes,
which is naturally accompanied by a lower transversal mobility
of the stripes.
It should also be noticed, that static disorder effects due to pinning
of the stripes should comparatively be stronger in ED materials, where
the stripes are expected to be less mobile. 

The formation of ordered arrays of stripes is
naturally accompanied by an (incommensurate) spin response,
which appears in neutron scattering at momenta displaced
by an amount $\varepsilon$ from the AF wavevector
$Q_{AF}=(\pm \pi/a, \pm \pi/a)$. While this effect has indeed been
observed in metallic HD materials, with $\varepsilon \sim x$ for $x$ up to 1/8
and then saturating ($\varepsilon \sim constant$) \cite{nota18}, 
comparatively detailed neutron 
scattering data are not available for ED systems. 
Our findings suggest that the incommensurate momenta should scale as
$\varepsilon \sim x/\sqrt{2}$. Moreover, while in HD materials
the spin excitation peaks appear in the $(1,0)$ and $(0,1)$
directions consistently with the vertical ordering of the stripes,
in ED materials the corresponding peaks should be observed 
in the $(1,1)$ and $(1,-1)$ directions, according to our
finding of diagonal stripe formation. 
However, the substantial disorder effects due to pinning could broaden
the spin peaks making the observation of incommensurability
more difficult. In this respect more local probes
like the NQR (successfully applied in the HD materials\cite{IMAI})
can turn out to be very useful.

As far as transport is concerned, no theory is presently available 
in the case of scattering due to (dynamically fluctuating)
stripes. Therefore no definite statement can be put forward
beside the following generic remarks. The first generic observation 
is that  the narrower charge distribution of stripes in 
ED systems, indicates a smaller transversal mobility.
This, together with the larger separation of the filled ED stripes 
accounts for the more extended insulating region of the ED systems with 
respect to the HD ones. Specifically this
could explain why the antiferromagnetic phase
is quite more robust upon electron doping  than upon hole doping
stressing the analogy between the stripes in ED systems and in nickelates.

As far as the longitudinal (i.e. along the stripes) metallicity
is concerned, it can be noticed that HFVS should be more
metallic than FDS. This is expected because in the longitudinal direction
HFVS are similar to an array of quarter-filled onedimensional wires.
If these wires escape the longitudinal CDW commensurate instability, 
this 1D structures should naturally be more metallic than 1D chains
with an empty or a full band \cite{notametallic}. 
If this is the case, a mechanism for
superconductivity based on coupled metallic chains \cite{EMERY1}
seems hardly applicable to the ED case at least at the low 
and intermediate doping regimes where (according to our calculations)
the stripes are expected to be less mobile in the transverse direction,
nearly insulating in the longitudinal direction, and far one from the
other. Of course this objection no longer holds when, by increasing
doping, the metallic character of the system increases and our
calculational scheme is no longer applicable. In this regime
the stripes will be closer and likely far from the completely
filled condition. In this latter doping
region, however, the proximity to the quantum critical point
observed near $x=0.17$ \cite{FOURNIER} opens the way to 
alternative mechanisms \cite{CDG} based on the scattering from
critical charge fluctuations.

The metallic properties of ED stripes should also be strongly
affected by disorder effects, which, as already mentioned above,
are expected to be more important when the stripes are less mobile.
In this regard, the recent observation \cite{ONOSE} that small amounts of
interstitial apical oxygen suppress superconductivity (but not metallicity)
giving rise to pseudogap effects is of obvious relevance. A
natural speculation could be that the interstitial oxygen triggers
a pinning of the stripes. The suppression of stripe {\em fluctuations}
then induces the suppression of superconductivity giving rise to
a metallic (non-superconducting) state with disordered static
(and locally well-formed) stripes with the consequent stronger reduction
of density of states near the Fermi level.

Another generic consequence is that the diagonal stripes of the ED
systems, 
should display rather different commensurability effects 
than in the HD materials. This is a consequence
of the different spacing and ordering direction as well as of
the different $T'$ crystal structure of the ED cuprates. 
In particular, in the ED case, the 1/8 
effects are not expected to be marked \cite{nota18b}
since the stripes are quite far apart and the collective
nature of the ordered state is likely affected by disorder
pinning.

In conclusion, the occurrence of stripes in a rather
realistic model for ED cuprates is a quite intriguing finding. 
According to our results,
the specific features of the stripes in
ED material are different from the features of stripes in HD
ones. Therefore, if the occurrence of stripes in electron-doped
systems will be experimentally confirmed, the comparison with the stripes
already observed in the hole-doped materials will likely 
shed light on the physical mechanisms
acting in both ED and HD cuprates. 
This calls for a search of experimental 
signatures and characterization of stripes in ED cuprates,
hopefully leading to a better understanding of the cuprates.

{\em Acknowledgments} We gratefully acknowledge stimulating discussions
with C. Castellani and C. Di Castro.  The interaction with G. Seibold
has been interesting and technically very helpful.

%%%%%%%%%%%%%%%%%%%%%%%%%%%%%%%%%%%%%%%%%%%%%%%%%%%%%%%%%%%%%%%%%%%%%%

\end{multicols}

\begin{references}
\bibitem{nota18} In particular, at $x=1/8$ this is consistent with HFVS
	separating antiphase spin domains with a periodicity $8a$
	\cite{TRAN}.
\bibitem{notametallic} Moreover, the effective intrastripe hopping
	is larger in the vertical (0,1) than in the diagonal (1,1)
	direction.
\bibitem{nota18b} Nevertheless, a careful inspection of Fig. 3
in Ref. \onlinecite{FOURNIER}, seem to indicate that a slight
	enhancement of the metal-insulator crossover temperature
	takes place for the sample at $x=0.13\approx 1/8$.
\end{references}
\end{document}